\documentclass[final,5p,times,twocolumn,authoryear]{elsarticle}
\usepackage{amsmath}
\usepackage{graphicx}

\journal{Journal of the Mechanics and Physics of Solids}
\begin{document}
\begin{frontmatter}
\title{Microstructural model for cyclic hardening in F-actin networks crosslinked by $\alpha$-actinin}

\author[hlm,jfr]{Horacio L\'opez-Men\'endez\corref{cor1}}
\ead{horacio.lopez-menendez@ijm.fr / horacio.lopez.menendez@gmail.com}
\author[jfr]{Jos\'e F\'elix Rodr\'iguez}
\ead{jfrodrig@unizar.es}
\cortext[cor1]{Principal corresponding author}
\address[hlm]{Cell Adhesion and Mechanics, Institut Jacques Monod (IJM), CNRS UMR 7592 \& Universit\`e Paris Diderot, Paris, France}
\address[jfr]{Aragon Institute for Engineering Research (I3A), University of Zaragoza, 50018 Zaragoza, Spain}

\begin{abstract}

The rheology of F-actin networks has attracted a great attention during the last years. In order to gain a complete understanding of the rheological properties of these novel materials, it is necessary the study in a large deformations regime to alter their internal structure. In this sense, \cite{Schmoller2010} showed that the reconstituted networks of F-actin crosslinked with $\alpha$-actinin unexpectedly harden when they are subjected to a cyclical shear. This observation contradicts the expected Mullins effect observed in most soft materials, such as rubber and living tissues, where a pronounced softening is observed when they are cyclically deformed.  

We think that the key to understand this stunning effect is the gelation process. To define it, the most relevant constituents are the chemical crosslinks -$\alpha$-actinin-, the physical crosslinks -introduced by the entanglement of the semiflexible network- and the interaction between them. As a consequence of this interaction, a pre-stressed network emerges and introduces a feedback effect, where the pre-stress also regulates the adhesion energy of the $\alpha$-actinin, setting the structure in a metastable reference configuration. Therefore, the external loads and the evolvement of the trapped stress drive the microstructural changes during the cyclic loading protocol.
In this work, we propose a micromechanical model into the framework of nonlinear continuum mechanics. The mechanics of the F-actin filaments is modeled using the wormlike chain model for semiflexible filaments and the gelation process is modeled as mesoscale dynamics for the $\alpha$-actinin and physical crosslink.The model has been validated with reported experimental results.
 
\end{abstract}

\end{frontmatter}

\section{Introduction}
Bio-polymeric meshworks has attracted a great attention as bio-materials because of their soft and wet nature, similar to many biological scaffolding structures. However, any given application requires a combination of mechanical properties, including stiffness, strength, toughness, damping, self-healing and fatigue resistance. The study of these structures can contribute to a better understanding of this new micro/nano technology and the cytoskeleton like structural building blocks, as was shown by \cite{keber2014,Gardel2004,Lieleg2010,Lieleg2011,Schmoller2008,Schmoller2009,Schmoller2010}. In order to improve the understanding of the rheological properties of these novel materials, it is necessary the study of extreme situations.  Nonlinear deformations can irreversibly alter the mechanical properties of materials. Traditionally, the actin networks have been considered as a system in a thermodynamic equilibrium which only could be driven out of equilibrium under the action of actin-myosin molecular motors \citep{Mizuno2008}.

Nevertheless, \cite{Schmoller2010} showed that the reconstituted networks of F-actin crosslinked with $\alpha$-actinin harden when they are subjected to cyclical shear. This observation contradicts the expected Mullins effect observed in most soft materials, such as rubber and living tissues, which soften by means of a cyclically deformation protocol \citep{diani2009}. Several mathematical models have been developed to describe the effect of stiffening and softening observed for cyclic loading protocols, but nobody has developed a model to describe the cyclic hardening effect reported by \cite{Schmoller2010}. In the following, we will summarize some of these approaches already performed for biopolymeric networks and at rubber and soft-tissue level.
 
\cite{Wolff2010,Wolff2012} developed a formalism called iGWLC $($inelastic glassy wormlike chain$)$ to link the nonlinear mechanical description of the wormlike chain model with the dynamic of the crosslinks. This model describes the experimental results of the observed effect of stiffening and softening, considering the crosslinks dynamics. This approach was done in the Fourier domain, which is suitable for the experimental description of the rheological experiments. Others models developed by \cite{van2008,Kim2009a,Kim2009b,Abhilash2012,Cyron2013} using computational techniques based on the reconstruction of the networks by means of different methods such as the Brownian dynamics and the finite element. The interaction between filaments is described by the transient-dynamics of crosslinks. These models are computationally demanding in order to gain information about the rheological properties.

At a tissue and rubber scale, several models have been developed for the standard Mullins effect. Usually, the description of the constitutive behaviour of this type of material relies on the identification of an appropriate strain-energy density function (SEF) from which stress-strain relations and local elasticity tensors can be derived \citep{Holzapfel2000}. A number of SEF has been proposed to describe the behaviour of soft tissue with damage. These models are based on the introduction of internal variables that account for non-physiological loading that drives soft tissue to damage. In most models, the main damage mechanism is associated with tear or plastic deformation of fibers.  \cite{Hurschler1997} proposed a micromechanical model for ligament behaviour that includes fiber failure. Similarly, we find the model of \cite{Arnoux2002} and \cite{Schechtman2002} for ligaments and tendons, or the work of \cite{Hokanson1997} for damaging arteries. \cite{Gasser2002} proposed a rate-independent multisurface elastoplastic constitutive model for soft tissue which introduced inelastic deformation of the collagenous component of the tissue. \cite{balzani2006}  proposed a discontinuous damage model for arteries in which the damage of the fibers is treated following classical continuous damage theory. Also, \cite{JFRodriguez} developed a constitutive model which accounts for different damage processes for matrix and fibers. Fibrous part was assumed to follow the wormlike chain model~\citep{Mackintosh1995} where damage was incorporated through the statistical distribution of the deformation at the fully extended length of collagen fiber bundles (crosslinks rupture). More recently, \cite{Saez2012} have proposed a microsphere based model for modelling damage in fibrous tissue where the directional statistics is used to describe the orientation of collagen fibbers within the tissue.

In this work, we propose a mathematical model within the framework of nonlinear continuum mechanics to describe the cyclic hardening effect reported by \cite{Schmoller2010}. We think that the key to understand this stunning effect is the gelation process in this kind of networks, where the most relevant constituents are the chemical crosslinks, defined by the $\alpha$-actinin, and the physical crosslinks, introduced by the entanglement of the semiflexible F-actin network. As a consequence of this interaction, a pre-stressed network emerges ~\citep{Schmoller2010,Lieleg2011} and introduces a feedback effect, where the pre-stress regulates the adhesion energy of the $\alpha$-actinin and sets the structure in a metastable reference configuration. Therefore, the external loads and the evolvement of the trapped stress drive the microstructural changes during the cyclic loading protocol.

We take as starting point the mechanics of networks with rigid crosslinks, using the wormlike chain model in the form proposed by~\cite{Mackintosh1995} and further developed by \cite{Palmer2008}, whereas the network is described using an homogenized continuum framework based on the eight chain network \citep{Arruda1993,Bertoldi2007,Palmer2008,Brown2009}. Afterwards, we introduce the inelastic effect as alterations in the contour length of the F-actin network. To define the phenomenological law that drives the changes in the contour length, we propose a simple model for the gelation process of the network based on the interactions between the physical and chemical crosslinkers.In the results section, we describe the monotonic and cyclic experiments, showing an opposite to the expected Mullins effect. The study of the evolution of the free parameters during the cyclic process give to us some ideas regarding the evolution of the microstructure through the experiment.

\section{Materials and methods}
\subsection{Entropic bundle network elasticity}

\begin{figure}[h]
\begin{center}
\includegraphics[width=8cm,keepaspectratio]{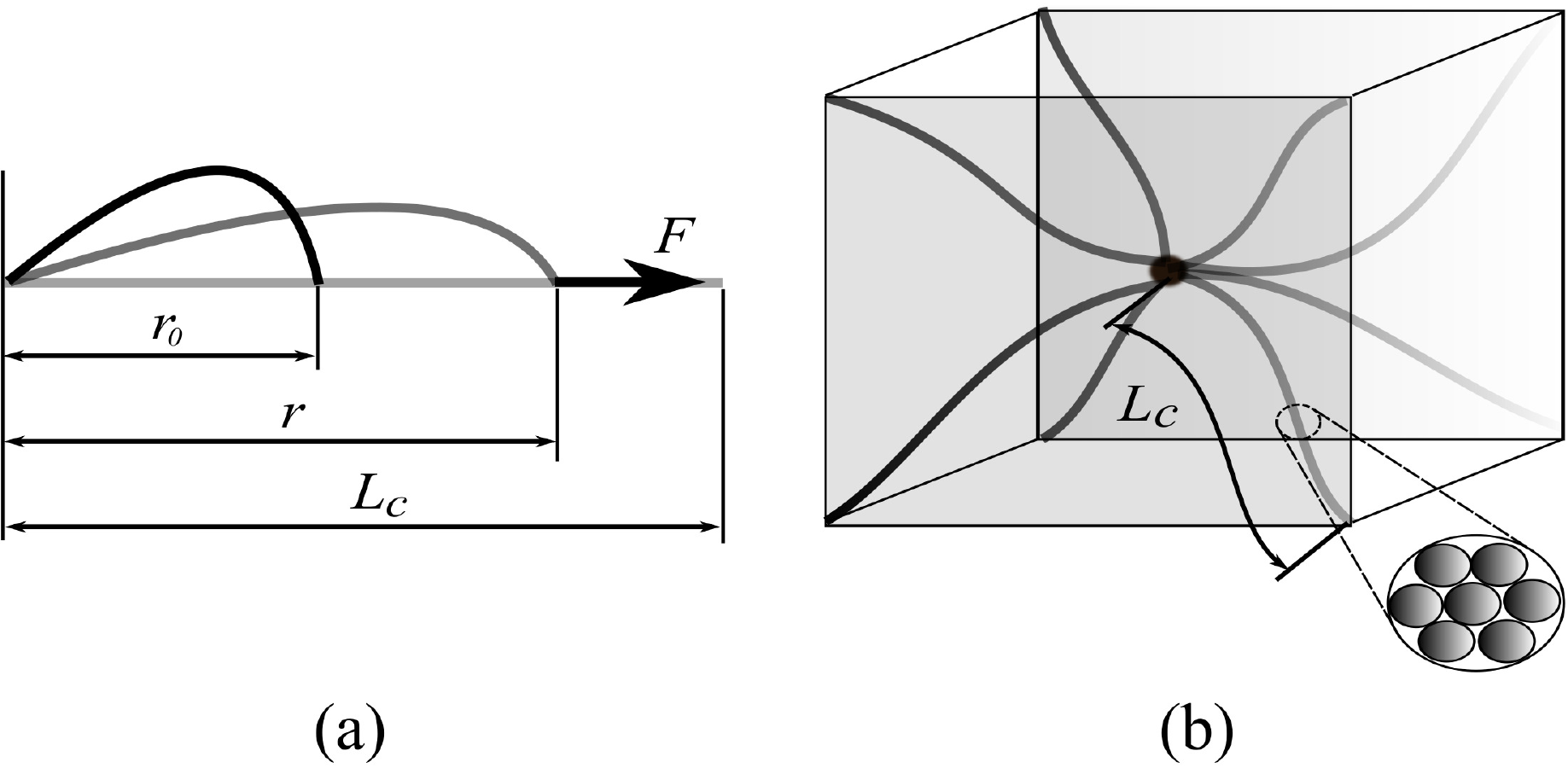}
\caption{a) Single filament schematic and b) Idealized eight chain model of an F-Actin network. }\label{Fig2}
\end{center}
\end{figure}

The mechanical behaviour of single actin filaments is governed by the worm-like chain $(wlc)$ model for semiflexible filaments, as proposed by~\cite{Mackintosh1995} to describe crosslinked polymer networks in which the force-stretch relationship is given by 
\begin{equation}\label{eq_1}
F_{wlc} =\frac{k_{B}T}{l_{p}}\left[\frac{1}{4\left(1-\displaystyle{\frac{r}{L_{c}}}\right)^{2}}\right]\left[\frac{\displaystyle{\frac{L_{c}}{l_{p}}}-6\left(1-\frac{r}{L_{c}}\right)}{\displaystyle{\frac{L_{c}}{l_{p}}}-2\left(1-\frac{r}{L_{c}}\right)}\right],
\end{equation}
where $k_{B}$ is the Boltzmann constant, $T$ is the absolute temperature, $L_{c}$ is the bundle length, $l_{p}$ is the persistence length, and $r$ represents the end-to-end distance (see Fig.~\ref{Fig2}a). The persistence length is defined as the length at which the entropic contributions to elasticity become important, as the bundle shows significant bending purely due to its thermal energy. A bundle with $L_{c}>>l_{p}$ bends, even without application of forces. 
In order to extend the model in Eq.~(\ref{eq_1}) from a single filament to a continuum description 
of the F-actin network, we adopt the approach proposed by~\cite{Palmer2008} based on the eight-chain model. In this model, the network is considered isotropic, and is idealized as a unit cube with eight chains, or bundles, extending from the center to each of the vertices of the cube (see Fig.~\ref{Fig2}b). The reference (undeformed) end-to-end distance of each bundle is $r_{0}$, so that $r_{0}=\sqrt{3}/2$. Due to applied stress the unit cube becomes a cuboid in the deformed configuration. If the unit cube is aligned with the principal stretch directions, it can be shown that the stretch of any chain in the unit cube, $\lambda$, is given by~\cite{Palmer2008}
\begin{equation}\label{eq_2}
\lambda=\frac{r}{r_0} = \sqrt{I_1/3}
\end{equation}
where $I_1$ is the first invariant of the right Cauchy-Green deformation tensor $\mathbf{C}=\mathbf{F}^T\mathbf{F}$, and $\mathbf{F}=\partial \mathbf{x}/\partial \mathbf{X}$ is the deformation gradient, where $\mathbf{x}$ is the position of a material point in the current configuration and $\mathbf{X}$ is the original position. Therefore, the end-to-end distance $r$ can now be written as $r=r_{0}\sqrt{I_1/3}$, and the force stretched relation in Eq.~(\ref{eq_1}) expressed in terms of the deformation tensor $\mathbf{C}$. 

From a continuum mechanics point of view, it is convenient to identify a strain energy density function for the network. This can be achieved by calculating the work done by each chain (integrating the filament force-extension expression in Eq. (\ref{eq_1})) and then multiplying the resulting expression by the filament density, $n$ (number of filaments per unit volume). Following the procedure proposed in \cite{Palmer2008} we obtain
\begin{equation}\label{eq_3}
\begin{split}
\psi_{wlc} &=\frac{n k_{B}T}{l_{p}}\left[\frac{L_c}{4\left(1-\displaystyle{\frac{r}{L_{c}}}\right)} -l_p\log \frac{L_c^2-2l_pL_c+2l_pr}{r-L_c} -c\right]
\end{split}
\end{equation}
where $c$ is a constant equal to the initial strain energy density from the filaments.
Since the F-Actin network is embedded in a nearly-incompressible fluid, the strain energy function of the network, $\Psi_{wlc}$, is rewritten as
\begin{equation}\label{eq_4}
\Psi_{wlc}(\mathbf{C},l_{p},L_{c})=  \psi_{wlc}(r,l_{p},L_{c})+p(J-1),
\end{equation}
where, $\Psi_{wlc}$ is defined for $J=\det \mathbf{F}= 1$. The scalar $p$ is an indeterminate Lagrange multiplier which can be identified as a hydrostatic pressure, and that is obtained from the equilibrium equations and boundary conditions. 

Using standard procedures from Continuum Mechanics, the Cauchy stress, $\boldsymbol{\sigma}$, can be derived from direct differentiation of Eq.~(\ref{eq_4}) with respect to $\mathbf{C}$~\citep{Holzapfel2000}
\begin{equation}\label{eq_5}
\begin{split}
\boldsymbol{\sigma}&=\frac{2}{J}\mathbf{F}\frac{\partial\Psi_{wlc}}{\partial \mathbf{C}}\mathbf{F}^T\\
&=\frac{nk_{B}T}{3l_{p}}\frac{r_{0}}{\lambda}\left[\frac{1}{4\left(1-\displaystyle{\frac{\lambda r_{0}}{L_{c}}}\right)^{2}}\right]\left[\frac{\displaystyle{\frac{L_{c}}{l_{p}}}-6\left(1-\frac{\lambda r_{0}}{L_{c}}\right)}{\displaystyle{\frac{L_{c}}{l_{p}}}-2\left(1-\frac{\lambda r_{0}}{L_{c}}\right)}\right]\mathbf{b}+p\mathbf{I}
\end{split}
\end{equation}
where $\mathbf{b}=\mathbf{F}\mathbf{F}^T$ is the left Cauchy-Green deformation tensor, and $\mathbf{I}$ is the second order identity tensor.

As mentioned before, the F-actin network is assumed to be an out of equilibrium
network. If we consider the effects of the kinetically trapped stress
over the network structure, we should consider that the end-to-end
distance in the reference configuration does not correspond to the distance
for zero force. For the Mackintosh model, Eq.~(\ref{eq_1}), the expression for the end-to-end
distance at zero force is $r_{0_{F=0}}=L_{c}\left(1-\frac{L_{c}}{6l_{p}}\right)$.
In order to describe the prestressed network we introduce an internal
variable $\epsilon$ which represents the degree of prestrain as a
fraction of $r_{0_{F=0}}$, as proposed by~\cite{Palmer2008}. Therefore the expression for $r$ becomes: 
\begin{equation}\label{rEq}
r=\lambda\left(1+\epsilon\right)L_{c}\left(1-\frac{L_{c}}{6l_{p}}\right)
\end{equation}
 
\subsection{Gelation process}
 
\begin{figure}[h!]
\includegraphics[width=8cm,height=8cm,keepaspectratio]{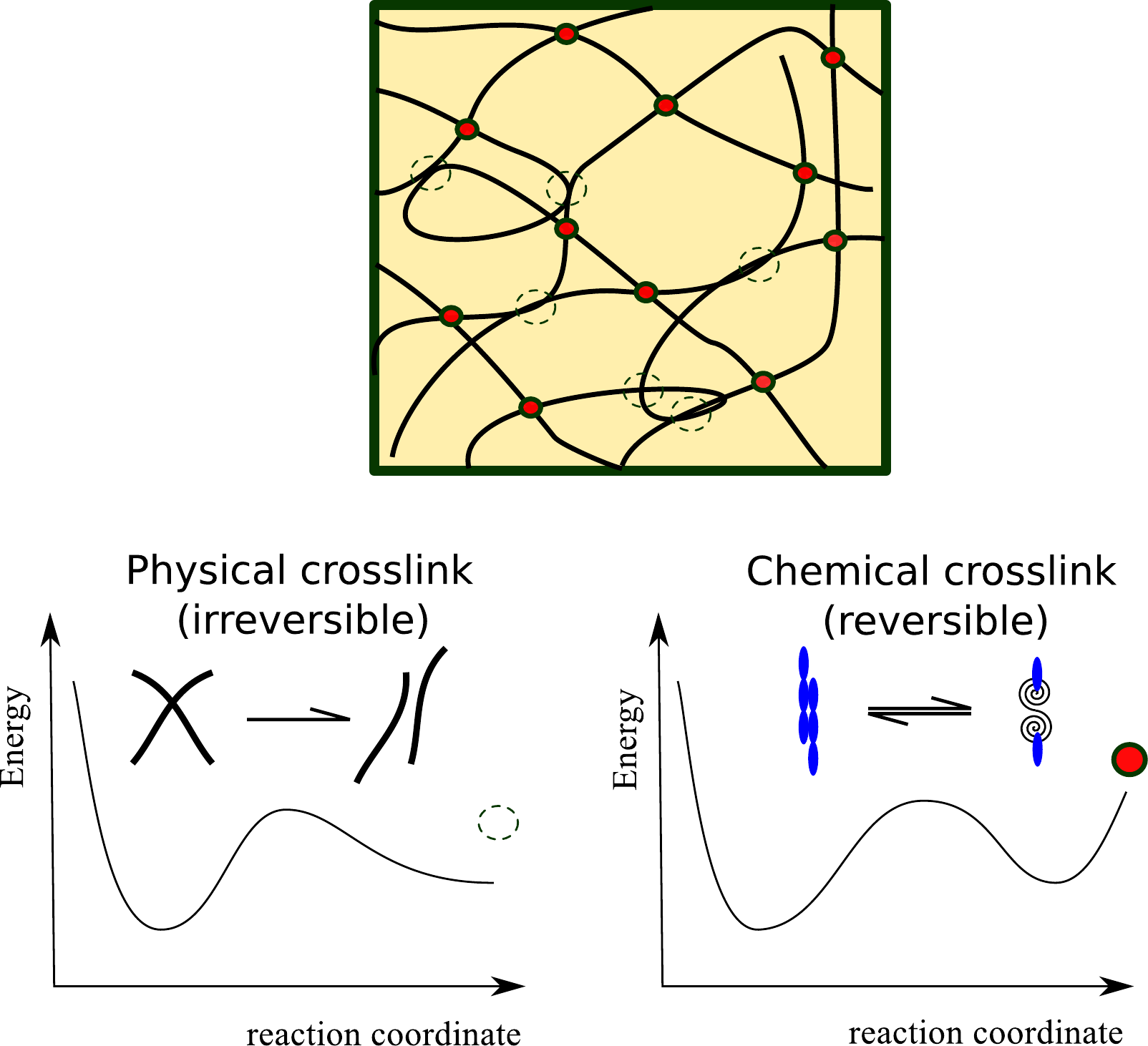}\centering
\caption{A simplified picture of the network is made by the combination between semi-flexible crosslinks, chemical crosslinks, and physical crosslinks. 
The chemical crosslinks are given by the $\alpha$-actin, which develop a reversible interaction; and the physical crosslinks, due to the entanglement between filaments, show an irreversible energy gap.}\label{gelation}
\end{figure}

The experiment performed by Schmoller represents a network with an intricate gelation process, where the rheological response is strongly dependent upon the conditions of the preparation \cite{witten2010}. The joint interaction between $\alpha$-actinin, dense fraction of semi-flexible filaments, polymerization, branching, fluctuations, and entanglement freeze the state of the network in a highly pre-stressed condition and generally in a metastable equilibrium~\citep{Lieleg2011,Lieleg2009}.  Figure \ref{gelation} shows a simplified scheme of the physical and chemical crosslinks and the interaction between them with the semiflexible network structure. For the general description of the gelation process we follow similar arguments as those exposed in classical polymers physics bibliography as \cite{deGennes1979,witten2010}.

\subsubsection{Chemical crosslinks}

The chemical crosslinks are given by the $\alpha$-actinin and if they are stable (for the stress and the time scales of the experiments), they provide a strong gelation process. On the contrary, if they are not stable but associated with a reaction that can proceed in both directions (as binding-unbinding of the crosslink), we speak of a weak gelation process and we expect to find some of the intricacies of glass transitions \citep{witten2010,deGennes1979}. In order to describe mathematically the chemical crosslinks, they can be modelled as a reversible two-state equilibrium process \citep{Brown2009,purohit2011}; as can be seen in the following expression: 

\begin{equation}
\frac { P_{ ub} }{ P_{ b} } =\exp -\frac {( \Delta G_{ 0 } - w_{ ext} )}{ k_{ B }T }  \label{eq_7}
\end{equation}

Where $P_{ub}$ defines the unbinding probability encompassing the states of unbinding, unfolding or flexible cross-link, and $P_b$ the binding probability encompassing the states of binding, folding or rigid cross-link. Since only these two states are possible, then $P_{ub} + P_b = 1$. The two-state model has the binded $\alpha$-actinin as the preferred low free energy equilibrium state at zero force and the unbinded $\alpha$-actinin as the high free energy equilibrium state at zero force. $\Delta G_0$  represents the difference in the free-energy between these states and $w_{ext}$ represents the external mechanical work that induce the deformation of the crosslink. $k_{B}T$ represents the thermal energy. 

If we consider the conservation of probability, $P_{ub}$ becomes:

\begin{equation}
P_{ ub} =\frac{1}{1+\exp\left[\frac {(\Delta G_{ 0 } -w_{ ext })}{ k_{ B }T } \right]} \label{eq_8}
\end{equation}

\subsubsection{Physical crosslinks and sacrificial bonds}

Any physical process which favours the association between certain points on different chains may lead to gels. The entanglement effect may drive a number of phenomena such as the formation of helical structures, micro-crystals, loops, and also electrostatic, hydrophobic, dipole-dipole interactions \citep{deGennes1979}. This is not an equilibrium process, but it corresponds to the progressive freezing of a number of degrees of freedom of the bundles via sticky interactions as can be observed in Figure~\ref{gelation}. 

The physical crosslinks can create networks with different kinds of stiffness according to the degree of entanglement. In this work we consider that the physical crosslinks will develop the role of sacrificial bonds and hidden length as was proposed by \cite{fantner2005,buehler2007,ducrot2014}. The sacrificial bonds in our case are defined as physical crosslink that break themselves, in a fragile way, before the chemical crosslinks of $\alpha$-actinin were broken. The hidden length is defined as the part of the molecule that was constrained from stretching by the sacrificial bond. This mechanism contributes with the toughness of the network by means the relaxation of stress and with the increment of the average contour length. We consider that the fraction of energy released due to the fragile breakage of physical crosslinks, working as sacrificial bonds, is dependent of the maximum level of deformation exerted over the network.
In order to describe the probability of fracture for the physical crosslink, we propose an Arrhenius-like relation,  in a similar way as was proposed by \cite {Bell1978,Evans2001,Bertoldi2007,buehler2007,Ciarletta2008}, but our phenomenological description employs the bundle stretch as a driving variable. In this model, the probability of failure is: 
\begin{equation}\label{EqPf}
P_{f}=P_{f_0}\exp\left[\kappa^{f}\left(\lambda_{\max}-\lambda^{f}_0\right)\right],
\end{equation}
where $\kappa^{f}, \lambda^{f}_0$ are a mesoscopic
material parameter associated with the activation energy needed to break the 
bond; $\lambda_{\max}$ represents the maximum stretch achieved by the bundle and $P_{f_{0}}$ represents the irreversible bond rupture at $\lambda_{\max}=\lambda{f}_0$. 

\subsubsection{Interaction between physical and chemical crosslinks}

\begin{figure}[h!]
\includegraphics[width=8.5cm,height=8.5cm,keepaspectratio]{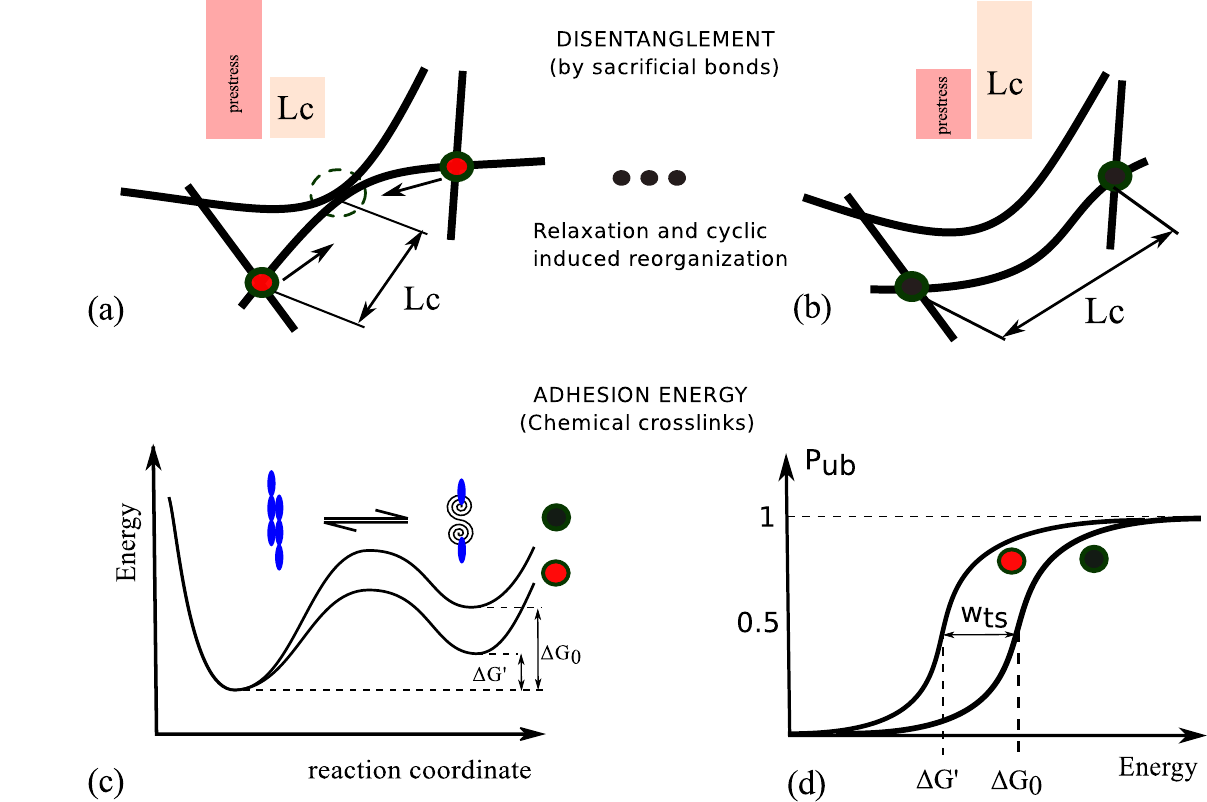}\centering
\caption{Interaction between physical and chemical crosslinks. (a) Gelation state in the reference configuration. The physical crosslinks induce prestrain over the bundles and reduce the contour length. The red dots represent a lower adhesion energy state of the $\alpha$-actinin. (b) Once the physical crosslinks, working as sacrificial bonds, release the energy, increase the contour length, reduce the prestrain over the $\alpha$-actinin and increase the adhesion energy. The black dots represent a higher adhesion energy of the $\alpha$-actinin. (c) Two energy landscape for the chemical crosslinks; with and without considering the effect of prestrain imposed the physical crosslinks. (d) Distribution function for the unbinding probability, $P_{ub}$ (Eq. \ref{eq_8} and Eq. \ref{eq_10}). The distribution without prestrain shows a higher transition point.}\label{interaction}
\end{figure}

During the gelation process, the physical crosslinks are created by the network entanglement. This process induces pre-stress across the network which is propagated through the bundles until the chemical crosslinks \citep{Lieleg2009,Lieleg2011}. The Figure \ref{interaction}.a illustrates this idea, where the interrupted line describes the physical crosslink, and the red dots represent the chemical crosslinks. In this configuration the pre-stress is higher and the contour length $(Lc)$ is lower due to the connectivity introduced by the entanglement. Therefore, the trapped stress into the structure is compensated by the deformation of the bundle and the chemical crosslinks. 
As a consequence, it is potentially able to induce conformational changes over the $\alpha$-actinin structure, as was described by \cite{Golji2009}.

Therefore, when the network is sheared at a certain critical strain, some physical crosslinks are more fragile-like and easy to break (those that behave as sacrificial bonds). Then, some of the trapped stress is relaxed and redistributed by means the disentanglement driven by the cyclic protocol. As can be observed in the Figure \ref{interaction}.b after the network reorganization, the pre-stress over the bundles and over the crosslinks is smaller, and the contour length $(Lc)$ is higher. 

The Figure \ref{interaction}.c describes the effect of the prestrain over the energy landscape of the chemical crosslinks. The red and the black dots mark the energy landscape with and without the conformational change introduced by the prestrain. Qualitatively is easy to see that the energy gap is lower under the action of prestrain over the $\alpha$-actinin, where the adhesion energy increases, changing from $\Delta G'$ to $\Delta G_0$, with $\Delta G_0 > \Delta G'$. In order to describe more quantitatively this fact, we modify the  Eq.\ref{eq_8} considering the role of the deformation energy exerted over the $\alpha$-actinin structure. This can be understood as a combined action of two mechanical regulation pathways over the $\alpha$-actinin reaction, where: $i)$ $w_{ext}$ represents the mechanical work induced by the macroscopic deformation that propagates through the network down to the $\alpha$-actinin. $ii)$ $w_{ts}$ represents the mechanical work introduced during the entanglement and the physical crosslinking formation which also deforms the $\alpha$-actinin structure. If we reorganize the terms defining $\Delta G'=\Delta G_{0}-w_{ts_i}$, the next expression can be obtained:

\begin{equation}
P_{ ub} =\frac{1}{1+\exp\left[\frac {(\Delta G'-w_{ ext })}{ k_{ B }T } \right]} \label{eq_10}
\end{equation}

The Figure \ref{interaction}.d illustrates the shifting effect due to the change introduced by $w_{ts}$. This tell us that the adhesion energy changes according to the state of the out-of-equilibrium forces into the network. Therefore, for the same macroscopic strain we observe different transition points, according to the internal prestrain of the structure, and as a consequence it explores different energy landscapes.

As we mentioned for the physical crosslinks, the experiments are in the mesoscale, where we are only able to measure macroscopic quantities as stress and strain. Since we are aiming to develop a constitutive model in the mesoscale, we propose the next phenomenological expression, using the previous expression as motivation:

\begin{equation}\label{eq_11}
P_{ub}=\frac{1}{1+\exp\left[\kappa^{ub}\left(\lambda^{ub}_{0}-\lambda\right)\right]},
\end{equation}

where the main driving force is $\lambda$, which is the average stretch over the bundle and is also proportional to the macroscopic shear strain. In order to simplify the mathematical treatment, we consider a linear relationship to approach the deformation energy of the $\alpha$-actinin crosslink as $\kappa^{ub}\lambda$. Also, $\kappa^{ub}\lambda_0$ is proportional to the intermediate adhesion energy $\Delta G'$. Then $\kappa^{ub}$ gives us an idea of the sharpness of the transition between states and $\lambda_0$ is the strain at which the transition is 0.5. If $\lambda_{0} << \lambda$, the network is easy to be remodeled showing a behaviour more fluid-like. If $\lambda_{0} >> \lambda$, the crosslink stability is higher and the probability of transition is very low. Consequently the network behaves as a solid-like structure. 

\subsubsection{Gelation and contour length }

\begin{figure}[h!]
\includegraphics[width=8cm,height=8cm,keepaspectratio]{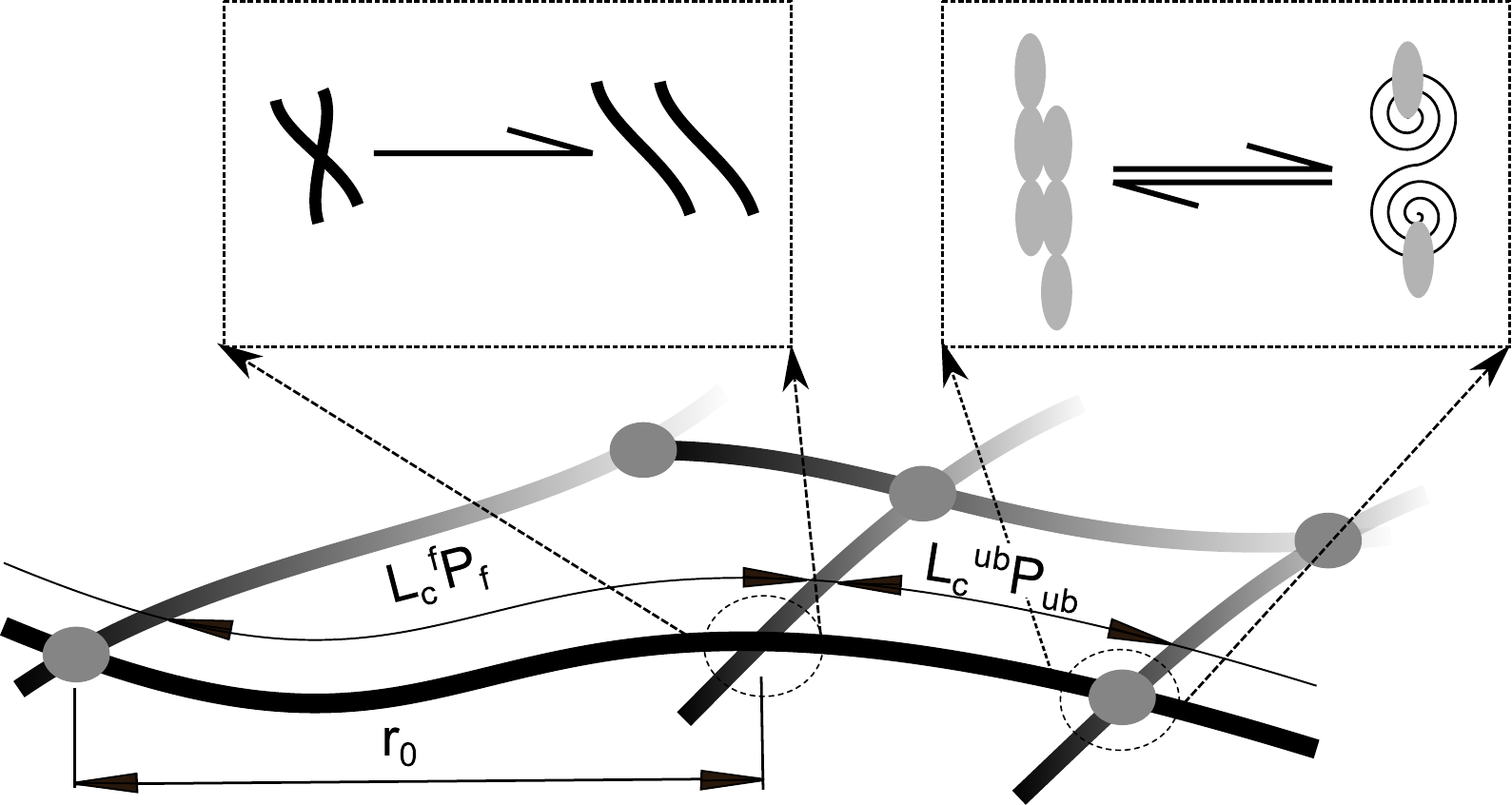}\centering
\caption{Semiflexible bundle structure and its interaction with the physical and chemical crosslink.}\label{Lc_bundle}
\end{figure}

Based on the previously described mechanism, we propose the ansatz, Eq.(\ref{EqLc}), for the average bundle length into the network. This can be considered as a stochastic variable dependent on the irreversible bound rupture probability (physical crosslink) and on the reversible unbinding probability (chemical crosslink). As can be observed schematically in the Figure \ref{Lc_bundle}.
\begin{equation}\label{EqLc}
L_{c}=L_{c}^{f}P_f+L_{c}^{ub}P_{ub},
\end{equation}
where parameters $L_{c}^{f}$ and $L_{c}^{ub}$ are regarded as material parameters determined from experiments.

\section{Results}

The proposed theory is used to describe the experiments conducted by \cite{Schmoller2010} on the artificially reconstituted F-actin networks cross\-linked with $\alpha$-actinin, where the network has an actin concentration of $c_a = 4.75$ $\mu$ M and a molar ratio of cross linking molecules to actin, $R = 1$, at $18$ $^\circ$C. For large enough concentrations of the cross linker, these networks show a pronounced nonlinear mechanical response to shear strain. These networks also show a profound network reorganization when are subjected to cyclic shearing. 

If we apply a monotonic and constant shear deformation rate, the network stiffening response starts at low levels of strain and continues almost linearly until reaching a maximum critical shear stress.  After that point, the stress decreases until reaching a plateau phase that slowly decreases towards zero as the shear strain increases. However, if the network is subjected to a cyclic shear strain $\gamma$ applied at a rate of $1.4\%$~s$^{-1}$ and always reaching the same maximum shear strain value, the network experiences a significantly different response from the first cycle. Schmoller et al. observed that, after the first loading cycle, each repetition of the deformation resulted in an increasingly larger linear regime, but also in a network that could withstand much higher stress. This result is in sharp contrast to the Mullins effect observed in rubber-like materials. However, this particular behaviour was also found to be very much dependent on the concentration of  crosslinks. 

In order to described the previous experiments, the model has been specialized for a pure shear experiment. Therefore, in terms of the shear deformation, $\gamma$, the bundle
stretch is given by $\lambda=\sqrt{1+\gamma^{2}/3}$. The
Cauchy shear stress-strain relationship and the remaining equations of the model are reduce to:

\begin{equation}\label{tau_wlc}
\tau=\frac{nk_{B}T}{3l_{p}}\frac{r_{0}}{\lambda}\left[\frac{1}{4\left(1-\displaystyle{\frac{\lambda r_{0}}{L_{c}}}\right)^{2}}\right]\left[\frac{\displaystyle{\frac{L_{c}}{l_{p}}}-6\left(1-\frac{\lambda r_{0}}{L_{c}}\right)}{\displaystyle{\frac{L_{c}}{l_{p}}}-2\left(1-\frac{\lambda r_{0}}{L_{c}}\right)}\right]\gamma, \\[2pt]
\end{equation}
\begin{equation}
r=\lambda r_0=\lambda\left(1+\epsilon_i\right)L_{c}\left(1-\frac{L_{c}}{6l_{p}}\right),\\[2pt]
\end{equation}
\begin{equation}
L_{c}=L_{c}^{f} \exp\left[\kappa^{f}\left(\lambda_{\max}-\lambda^{f}_0\right)\right]+
\frac{L_{c}^{ub}}{1+\exp\left[\kappa^{ub}_i\left(\lambda_{0_i}^{ub}-\lambda\right)\right]}.
\end{equation}

As can be seen, we arrive to a compact set of coupled equations where only three parameters, indicated with the subindex $i$, change during the cyclic experiment. 

\subsection{On the parameters of the model}

There are two kinds of parameters into the coupled set of equations:
At one side, the typical values for the semiflexible-$wlc$ model with rigid crosslinks as ($L_c,l_p,\epsilon,n$). The plausible values for the orders of magnitude can be easily found in the literature as in \cite {Gardel2004,Palmer2008,Lieleg2010}. More specifically, the density of actin filaments $n$, represents a proportionality factor and it was adopted from \cite{Palmer2008}. The persistence length $l_{p}$, was taken as $17.48\mu m$ \citep{Gardel2004}. We should point out that the persistence length is also dependent on the crosslink concentration and loading \citep{Gardel2004,Lieleg2010} and should be defined as a stochastic variable as well. However, in order to simplify the model, we consider $l_p$ as a constant parameter in the following. The contour length contributions ($L_{c}^{f},L_c^{ub}$)  were estimated in the range of values of $L_c$ described in the experiment of~\cite{Schmoller2010}.
 
 On the other side, according to this model, the parameters associated with the crosslink dynamic encode the transitions which induce remodelling into the network. 
The parameters ($\kappa_i^{ub}, \lambda_{0_i}^{ub}$), change during the cyclic experiment. These values represent an indirect measure of the adhesion energy of chemical crosslinks of $\alpha$-actinin. $\lambda_0^{ub}$ describes the transition point in the contour length of the network filament and $\kappa^{ub}$ the sharpness of this transition.
The parameters ($\kappa^f,\lambda_0^f$), don't evolve during the cyclic parameter because they depend on the maximum level of strain. They represent the failure dynamics of the physical crosslinks and their role is more dominant during the regime of large deformations. In order to avoid the re-stiffening and guarantee the network softening we should keep the relation $\lambda r_0 <<  Lc$.

\subsection{Simulation of the monotonic and cyclic loading experiments}

\begin{figure}[h!]
\includegraphics[width=8.5cm,height=8.5cm,keepaspectratio]{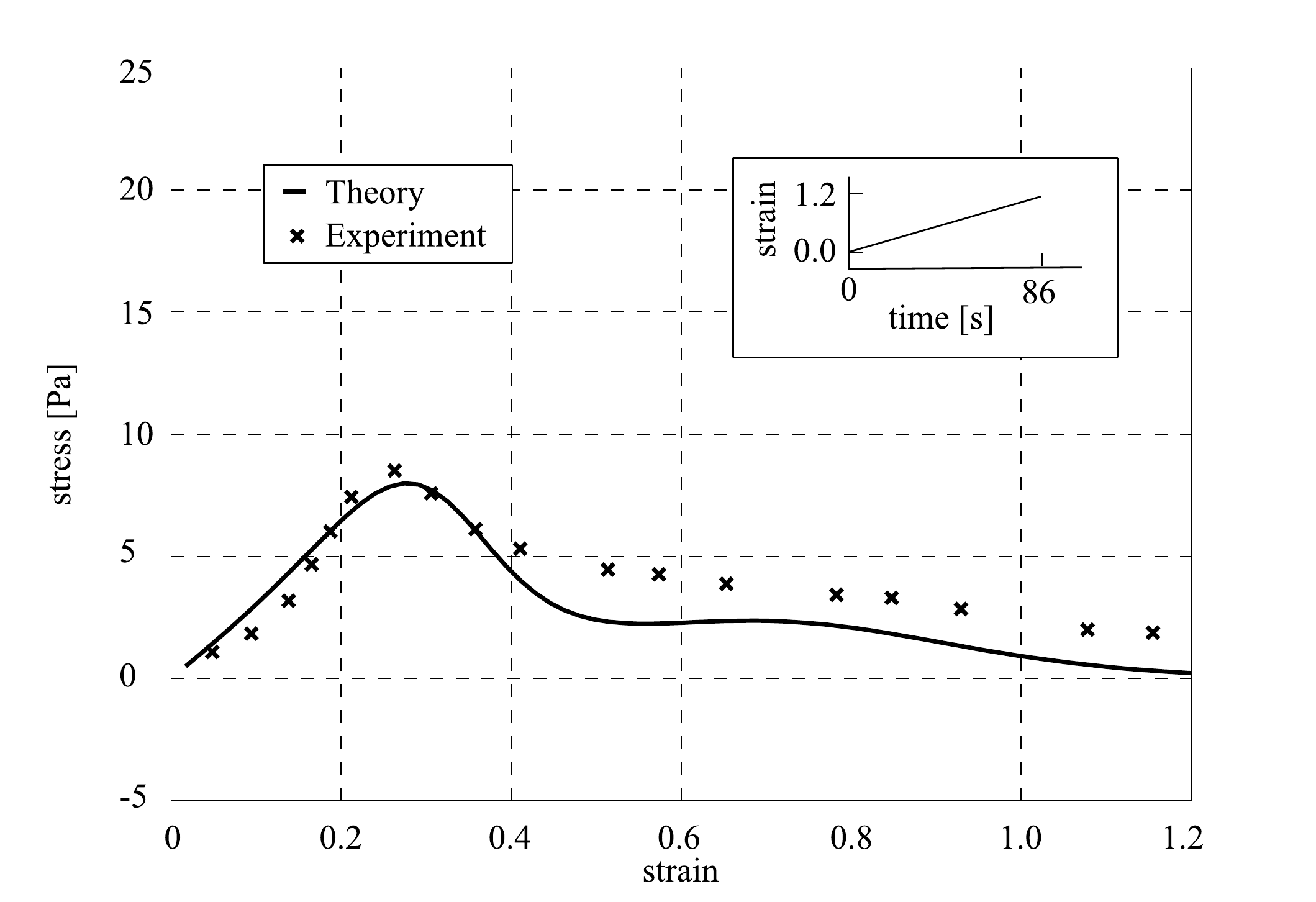}\centering
\caption{Monotonic shear experiment with maximum shear strain, $\gamma_{0}=1.2$. The figure shows the effect of softening of the network associated with the crosslinks
unbinding.}\label{monotonic}
\end{figure}

The monotonic experiment, see Figure \ref{monotonic}, shows the results of the model (solid line) along with the experimental data from~\cite{Schmoller2010} (black crosses) for a monotonic loading experiment in which the network has been sheared up to a maximum shear strain, $\gamma_0=1.2$, at a strain rate of $1.4\%$~s$^{-1}$ (see inset in Figure~\ref{monotonic}). The model predicts strain hardening response to start at $\gamma\approx0.1$ and continues until reaching a maximum shear stress $\tau_{max}\approx9Pa$ at $\gamma\approx0.28$ after which the stress decreases until reaching a plateau phase that slowly decreases towards zero as the shear strain increases. For strains lower than 0.4 the model closely follows the experimental data.  For larger shear strains, however, the model predictions are constantly biassed from the experimental data.

The identified model parameters obtained by best fitting of the experimental data are summarized in Table~\ref{tabR1} (see Apendix). The identified parameters are in good agreement with those found in the literature. \cite{Palmer2008} reported a boundle prestrain of 3\% with F-actin networks with lower actin/crosslink concentration ratio (R=0.03 and R=0.5). The initial contour length for our model (9 $\mu$m) is in good agreement with the mean mesh size of the network reported by \cite{Schmoller2010}.

\begin{figure}[h!]
\includegraphics[width=8cm,height=8cm,keepaspectratio]{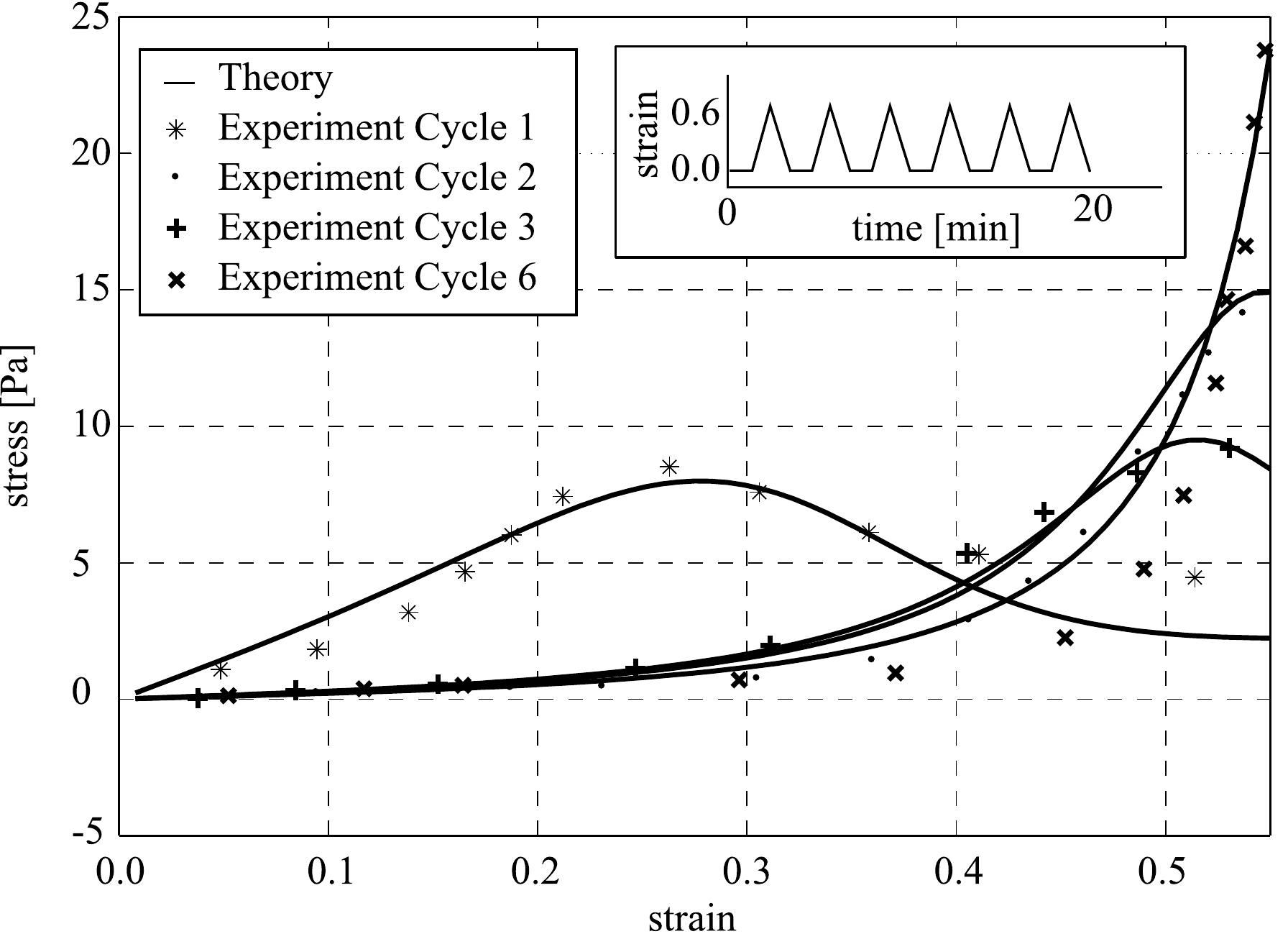}\centering
\caption{Cyclically sheared with $\gamma_{0}=0.55$. The figure shows the effect
of hardening of the network as the number of cycles increase}\label{cyclicloads}
\end{figure}

\begin{figure}[h!]
\includegraphics[width=8cm,height=8cm,keepaspectratio]{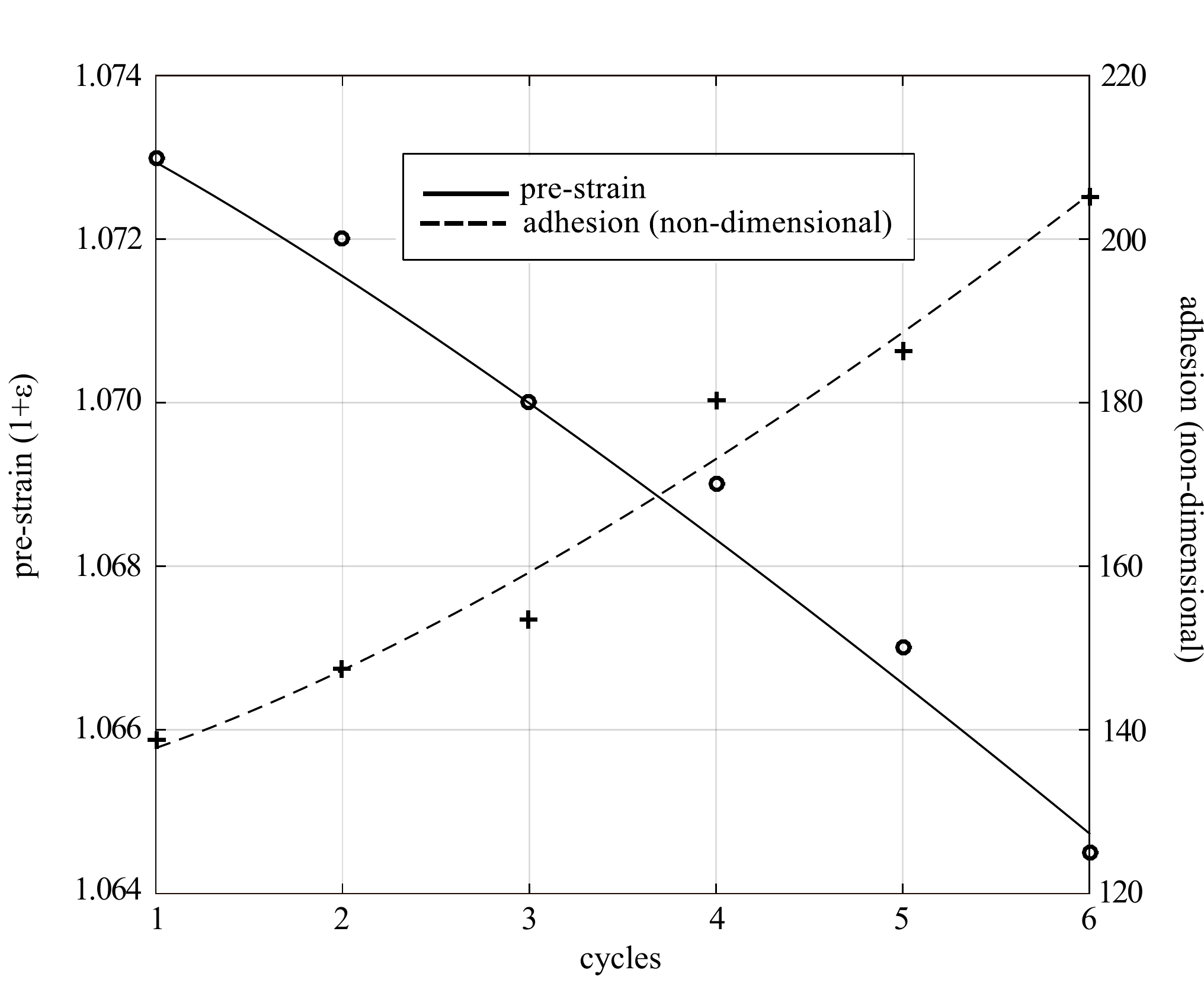}\centering
\caption{Evolution of the prestrain 1+$\epsilon$ and the mesoscale crosslinks adhesion $\kappa_i^{ub}\lambda_{0_i}^{ub}$}\label{prestress_adh}
\end{figure}

During the cyclic loading protocol, the response of the network is very different to the observed for monotonic loading protocol. As we can see in Figure \ref{cyclicloads}, after each strain cycle the linear regime gets larger whereas the network is able to withstand a higher maximum stress, in sharp contrast to the Mullins effect observed in rubber-like materials. When we apply the proposed model exploring the parametric space for $\kappa^{ub}_i$, $\lambda_{0_i}^{ub}$ and $\epsilon^i$, we find that it is able to fit quite well the experimental measurements of Schmoller, as can be observed in Figure~\ref{cyclicloads}. Evolving the three parameters for each cycles, we observe how the linear response of the network becomes larger, and how the network is able to reach a larger stress, as a network with rigid crosslinks (see Table~\ref{tabR2} in Apendix).

To understand the effects behind the different set of parameters used to fit the data, the Figure~\ref{prestress_adh} shows the relation between the bundle prestrain and the mesoscale approximation for the adhesion of crosslinks, $\kappa_i^{ub}\lambda_{0_i}^{ub}$. The figure demonstrates that the prestrain decreases monotonically with the number of cycles whereas the adhesion of crosslinks increases indicating the stabilization of the network. To describe it more quantitative, we fit the evolution of the parameters with the function $f=ax^b+c$, where the numerical values of the parameters for the prestrain are: $a=-1.06e-3; b=1.21; c=1.074$. And for  $\kappa_i^{ub}\lambda_{0_i}^{ub}$ are: $a=5.582; b=1.437; c=132.2$; showing a good fitting of the relationship between the parameters. This shows that the exploring parameters can fall in a master relation with almost the same scaling exponent $b$ for the bundle prestrain and for the stability of the chemical crosslinks.

\begin{figure}[h!] 
\centering
\includegraphics[width=8cm,keepaspectratio]{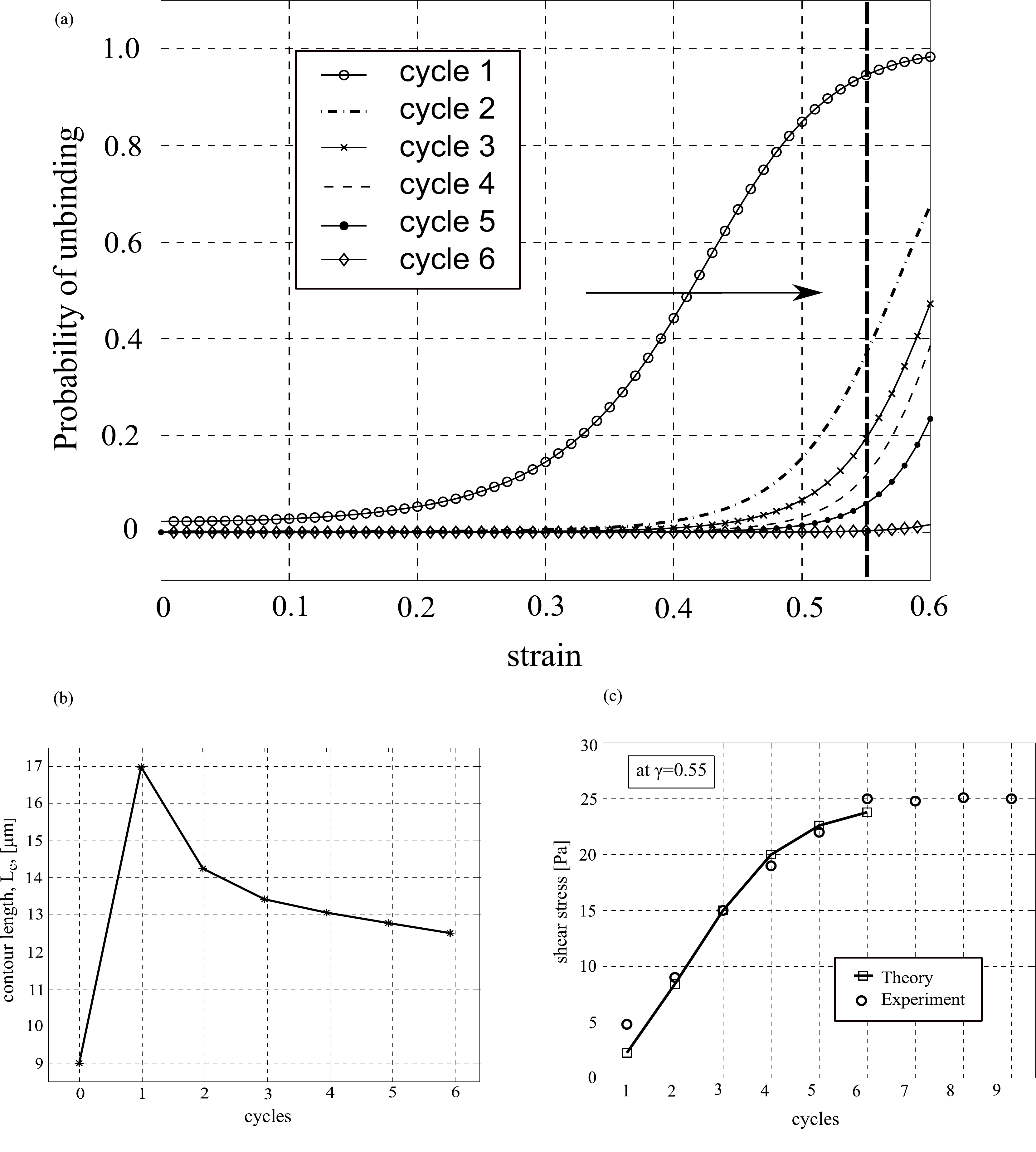}
\caption{$(a)$Dependence among the unbinding probability $P_{ub}$ and strain $\gamma$ for different number of cycles.$(b)$ Evolution of the contour length as a function of the number of cycles.$(c)$Difference between theoretical and experimental values of stress achieved at $\gamma=0.55$. 
\label{results}}
\end{figure}

This effect is in agreement with the proposed model explained in the Figure \ref{interaction} where the release of the trapped pre-stress drives the increase of the adhesion energy and, as a consequence, the transition point increases. The Figure~\ref{results}$.a$  plots the changes induced on  $P_{ub}^i$, after each load cycle (computed using Eq.~\ref{eq_11}). It is observed that, as the number of cycles increases the unbinding probability, $P_{ub}$, decreases for $\gamma=0.55$. The implications of this behaviour are firstly that the network structure stabilizes with a sufficient number of cycles and secondly, that the contribution with the contour length due to the reversible unbinding can be neglected.

Figure \ref{results}$.b$ shows the evolution of the contour length, $L_c$, as a function of the number of cycles. For the very first cycle, $L_c$ experiences the maximum increment due to the effect of the reversible and irreversible cross-links rupture. As the number of cycles increases the unbinding probability, $P_{ub}$ decreases as does the contour length due to the reformation of the reversible cross links, leading to an increment of the network stiffness in the large deformation range.

Figure \ref{results}$.c$ shows the effect of cyclic hardening for a $\gamma_{\max}=0.55$ on the maximum shear stress, $\tau$, reached at the maximal cyclically applied strain $\gamma_{\max}$. The figure illustrates that for the first cycles, the incremental rate in the reached shear stress is higher, evidencing more significant structural changes, but after some cycles the shear stress reaches a steady value. This evolution is also followed by significant changes in the shape of the stress-strain curve as shown in Figure \ref{cyclicloads}. In terms of model parameters, it implies that the network structure does not evolve with subsequent cycles. Therefore, a stable elastic response is obtained between different cycles.

\section{Discussion and Conclusions}

This work proposes a mathematical model to explain the experimental studies conducted on the reconstructed F-actin networks. The model is able to explain
the observed effects of softening, when the network is working in a regime of monotonic loading, and also the hardening, when the network is working under a cyclic loading protocol. The softening effect experienced by the networks is well documented for soft tissues-like materials and rubbers. On the contrary, the hardening induced by cyclic strain has not been observed in other rubber-like materials and seems to be associated with this type of structures.

The proposed model relies on the worm-like chain model for semiflexible filaments which depends on three structural network parameters, i.e., filament contour length, $L_c$, filament persistent length, $l_p$, and the undeformed end-to-end filament distance, $r_0$. One of the key parameters in the state of the network is the value of $r_0$, which is a function of $L_c$ (see Eq. \ref{rEq}). If $L_c$ increases $r_0$ decreases. The intuitive image for the $wlc$ model is that the closer the $r$ value to $L_c$, the response of the filament is more nonlinear and larger the tangent stiffness of the network. For bundles in networks with rigid crosslinks, the stress on the filament approaches to the locking point where the filament breaks at $r\approx L_c$ and the network collapses. In this regard, the proposed model introduces a split of the crosslinks dynamics as chemical (reversible) and physical (irreversible) crosslinks disruption that opens the possibilities of alternative dynamics which are able to reproduce the mechanical behaviour of the reconstituted F-actin networks observed in the experiments.

In this model we consider that the physical crosslinks impose a double effect on the network. On the one hand they prestrain the bundle, but on the other hand, the stretch of the chemical crosslinks of $\alpha$-actinin tilts the energy landscape of the crosslinks toward a state of less adhesion energy. During the cyclic experiment, a certain amount of physical crosslinks are broken and consequently, the prestrain over the bundles and crosslinks decreases. Therefore, the values of the $\Delta G'$ or $\kappa^{ub}\lambda_{0}^{ub}$ increase.  As a consequence, the probability of unbinding $P_{ub}$ decreases (see Figure~\ref{results}b) and the contour length $L_c$ of the network decreases as well (see Figure~\ref{results}c), showing an increase in the stiffening. The gelation state of the network changes from weak gelation and high bundle prestrain towards a stable state and with lower values of prestrain, which means a solid-like network. 

The development of a phenomenological model at mesoscale helps the characterisation of novel materials. Nevertheless, future works are needed to improve the estimations used in the dynamics of crosslinks in terms of the ratios between the concentration of F-actin, the concentration of $\alpha$-actin and the conditions of preparation of the network, in order to have an estimation of the tilted energy landscape of $\alpha$-actinin which represents a very difficult task.

 \cite{Yao2013} have reported dynamic nonlinearities in biopolymer F-actin networks crosslinked by $\alpha$-actinin-4. They observed that the applied stress delays the onset of relaxation and flows, enhancing gelation and extending the regime of solid-like behaviour to much lower frequencies. They suggest that this macroscopic network response can be accounted for at the single molecule level by the increased binding affinity of the crosslink under load, characteristics of the catch-bond-like behaviour ~\citep{Choi2005,Thomas2008b,Zocchi2009}. On the contrary, our approach explains the increase of adhesion energy at the network scale, by means the interaction between physical and chemical crosslinks and the relaxation of the trapped stress, due to the sacrificial bonds.
 
The model presented shows an alternative to extend the $wlc$ to describe the mechanical state of semiflexible networks with a more complex gelation process by considering the dynamics of the crosslinks. At the same time, the proposed mesoscale model, within the framework of continuum mechanics, can be easily incorporated to computational simulations based on the finite element method, in order to consider more complex geometries. 
The effect introduced by the cyclic shear lead us to speculate on the role of molecular motors of actin-myosin in the cytoskeleton. Molecular motors are capable of applying cyclic strain to the bundle structure, helping to modify the internal prestress of the crosslinks protein structures. In this regard, it seems that the role of $\alpha$-actinin into the cytoskeleton structure could be more complex than just a rigid cross-linker. Additional experimental studies are required to better understand the interaction between molecular motors, crosslinks, and actin filaments. However, the role of crosslinks dynamics should be considered in future developments of constitutive models for cytoskeleton-like structures.

\section*{Acknowledgements}
The authors wish to thank to the Dr. Kurt Schmoller and Prof. Dr. Andreas Bausch from the Technical University of Munich $($TUM$)$ for sharing the raw data and comments from its experiments. Also we wish to thank to the anonymous reviewer for they helpful suggestions that helped us to improve the manuscript.  H.L-M. thanks the Aragon's government and the University of Zaragoza for the FPI-DGA fellowship.

\section*{Apendix}
\begin{table}[h!]
\centering
\begin{tabular}{lll}
\hline
$1+\epsilon$ & Bundle prestrain & $1.0730$\tabularnewline
$n$&Density of actin filaments& $9.6e19$ [m$^{-3}$]\tabularnewline
$k_BT$& Thermal energy & $4.1$ [$p$N$n$m]\tabularnewline
$l_{p}$ & Persistence length & $17.48$ [$\mu$m]\tabularnewline
$L_{c}^{f}$ & Contour length & $10.5$ [$\mu$m]\tabularnewline
$L_{c}^{ub}$ & Contour length & $4.75$ [$\mu$m]\tabularnewline
$\lambda_{0}^{f}$ & Charact. stretch irrev. crosslinks & $1.024$\tabularnewline
$\lambda_{0}^{ub}$ & Charact. stretch rev. crosslinks & $1.028$\tabularnewline
$\kappa^{f}$ & Nondim. irreversible crosslinks stiffness & $7.1$\tabularnewline
$\kappa^{ub}$ & Nondim. reversible crosslinks stiffness & $135$\tabularnewline
\hline
\end{tabular}
\caption{Model parameters for the monotonic experiment}\label{tabR1}
\label{table1}
\end{table}

\begin{table}[h!]
\begin{tabular}{ccccc}
\hline
Cycle & $1+\epsilon_i$ &$\kappa_i^{ub}$ & $\lambda_{0_i}^{ub}$\\
\hline
1 & 1.0730 & 135 & 1.028 \\
2 & 1.0720 & 140 & 1.053 \\
3 & 1.0700 & 145 & 1.059 \\
4 & 1.0690 & 170 &  1.061\\
5 & 1.0670 & 175 & 1.065 \\
6 & 1.0645 & 190 & 1.080 \\
\hline 
\end{tabular}\centering\
\caption{Model parameters of the case of cyclic loading. Parameters $\kappa^{ub}_i$ and $\lambda_{0_i}^{ub}$ correspond to reversible cross linking occurring in the network under the action of cyclic loading. Parameter 1+$\epsilon$ refers to the prestrain into the network.}\label{tabR2}
\end{table}

\section*{References}

\bibliographystyle{unsrt}
\addcontentsline{toc}{section}{\refname}\bibliography{Lopez_Menendez_JMPS}

\end{document}